\documentclass[conf]{new-aiaa}

\usepackage[utf8]{inputenc}

\usepackage{flushend}
\usepackage{lipsum}
\usepackage{xspace}
\usepackage{todonotes}
\usepackage{soul}

\usepackage{amsmath}
\usepackage{amssymb}

\usepackage{booktabs}
\usepackage{array}

\usepackage{pifont}
\usepackage{float}
\usepackage{booktabs}
\usepackage{graphicx}
\usepackage[version=4]{mhchem}
\usepackage{subcaption}
\usepackage{siunitx}
\usepackage{longtable,tabularx}
\usepackage{wrapfig,lipsum,booktabs}
\usepackage[table]{xcolor}
\usepackage{ulem}
\definecolor{lightpink}{RGB}{255, 230, 230}
\definecolor{aquamarine}{rgb}{0.5, 1.0, 0.83}
\definecolor{applegreen}{rgb}{0.55, 0.71, 0.0}
\definecolor{babypink}{rgb}{0.96, 0.76, 0.76}
\usepackage{xcolor} % add this in your preamble

\definecolor{limegreen}{RGB}{50,205,50}
\definecolor{DodgerBlue}{RGB}{30, 144, 255}

\usepackage{xcolor}
\usepackage{soul}

\sethlcolor{yellow}

\newcommand{\KAL}{\textcolor[HTML]{FF8A80}{\texttt{KAL690}}}
\newcommand{\SIA}{\textcolor[HTML]{3F51B5}{\texttt{SIA608}}}
\newcommand{\AHK}{\textcolor[HTML]{EA4335}{\texttt{AHK128}}}
\newcommand{\ABL}{\textcolor[HTML]{AB47BC}{\texttt{ABL748}}}
\newcommand{\AAR}{\textcolor[HTML]{FF8A80}{\texttt{AAR762}}}

\soulregister\KAL0
\soulregister\SIA0
\soulregister\AHK0
\soulregister\ABL0
\soulregister\AAR0

\title{Probabilistic Multi-Agent Aircraft Landing Time Prediction}
\author{Kyungmin Kim\footnote{Graduate Student, Department of Advanced Air Transportation, 2kilomn@kau.kr, AIAA Student Member}, Seokbin Yoon\footnote{Research Assistant, Department of Air Transport, Transportation, and Logistics, sierra.bin@kau.ac.kr}, Keumjin Lee\footnote{Professor, Department of Air Transport, Transportation, and Logistics, keumjin.lee@kau.ac.kr, AIAA Member}}
\affil{Korea Aerospace University, Goyang, 10540, South Korea} 

\begin{document}

\maketitle

\begin{abstract}
Accurate and reliable aircraft landing time prediction is essential for effective resource allocation in air traffic management. However, the inherent uncertainty of aircraft trajectories and traffic flows poses significant challenges to both prediction accuracy and trustworthiness. Therefore, prediction models should not only provide point estimates of aircraft landing times but also the uncertainties associated with these predictions. Furthermore, aircraft trajectories are frequently influenced by the presence of nearby aircraft through air traffic control interventions such as radar vectoring. Consequently, landing time prediction models must account for multi-agent interactions in the airspace. In this work, we propose a probabilistic multi-agent aircraft landing time prediction framework that provides the landing times of multiple aircraft as distributions. We evaluate the proposed framework using an air traffic surveillance dataset collected from the terminal airspace of the Incheon International Airport in South Korea. The results demonstrate that the proposed model achieves higher prediction accuracy than the baselines and quantifies the associated uncertainties of its outcomes. In addition, the model uncovered underlying patterns in air traffic control through its attention scores, thereby enhancing explainability. Code is available at this repository: \href{https://github.com/atmcl/multiagent-landing-time-prediction}{\textcolor{DodgerBlue}{\texttt{https://github.com/atmcl/multiagent-landing-time-prediction.}}}
\end{abstract}

\section{Introduction} 
\lettrine{T}{he} rapid growth in global air travel demand has led to increased congestion and operational complexity around major airports~\cite{icao2024annual}. One of the key challenges arising from increased air traffic is the uncertainty of aircraft trajectories, which is driven by factors such as high traffic density and dynamic flow patterns. In particular, the trajectories of arriving aircraft are highly affected by various factors including weather and/or surrounding aircraft, resulting in increased uncertainty in their landing times. Such uncertainties undermine the efficiency of ground resource management and complicate the decision-making processes of airport operators and related stakeholders~\cite{heidt2016robust,ng2017robust}. 

To mitigate these uncertainties and provide reliable predictions, various aircraft trajectory models have been developed. Historically, physics-based approaches that rely on aircraft kinematic and aerodynamic performance data, such as the base of aircraft data (BADA)~\cite{nuic2010bada}, have served as the foundation for trajectory prediction~\cite{gallo2007trajectory,porretta2008performance}. However, a fundamental limitation of physics-based models is their inability to adequately incorporate external factors such as interventions by human air traffic controllers (ATCo)~\cite{hamed2013statistical}.

%very short term and guidance only

Motivated by these limitations, data-driven trajectory models have rapidly emerged as promising alternatives~\cite{de2013machine}. These models can capture intricate trajectory patterns that physics-based approaches often fail to represent by leveraging historical trajectory data. Numerous previous studies have employed classical regression techniques, such as linear regression, due to their simplicity~\cite{hong2015trajectory,hamed2013statistical}. Other types of models, such as Gaussian mixture models (GMMs), have also been explored to learn statistical patterns in aircraft trajectory datasets~\cite{barratt2018learning,paek2020route,jung2025inferring}. 

Recent advances in surveillance technologies, particularly automatic dependent surveillance–broadcast (ADS-B), have made large-scale trajectory data increasingly accessible, and the rapid progress in deep learning technologies has provided powerful tools for modeling complex patterns within these data. In one study, feed-forward neural networks (FFNs) comprising stacked linear layers have been employed for flight time prediction in the terminal airspace~\cite{wang2018hybrid,wang2020automated}. Long short-term memory networks (LSTMs) have been widely explored in other studies on trajectory prediction to capture the temporal characteristics of aircraft trajectories better~\cite{ma2020hybrid,shi20204,yoon2023improving,tran2022aircraft}.

Recently, Transformers have garnered significant attention~\cite{vaswani2017attention}, owing to their excellent performance in modeling complex, long-range temporal dependencies in trajectory data~\cite{tong2023long,silvestre2024multi,huang2025efficient}. Several studies have explored different trajectory representations, such as binary encoding~\cite{guo2024flightbert++} and inverted embedding~\cite{yoon2025aircraft}, to better adapt trajectory data to the Transformer architecture. However, these previous studies are limited because they do not consider the multi-agent nature of air traffic operations. Because aircraft trajectories are inherently influenced by surrounding traffic and are continuously controlled by ATCo for separation assurance, trajectory prediction models should incorporate these multi-agent interactions.

In response, some studies have explored multi-agent trajectory modeling to capture inter-aircraft interactions. Agent-aware attention, introduced in AgentFormer~\cite{yuan2021agentformer}, has been adapted to multi-agent trajectory modeling to learn the dynamic interactions between aircraft~\cite{choi2023multi}. Another study integrated learned multi-agent interactions with airport weather information to improve prediction accuracy~\cite{nguyen2025multi}. Although these studies have made progress in modeling multi-agent aircraft interactions, they do not account for the uncertainty in their predictions. 
Air traffic is inherently dynamic and influenced by various uncertain factors, such as weather and ATCo interventions; therefore, relying on a single deterministic estimate of the landing time of an aircraft could fail to capture the complexity and realities of actual operations.

\vspace{5pt}
\begin{figure}[h!]
    \centering
    \includegraphics[width=1.0\textwidth]{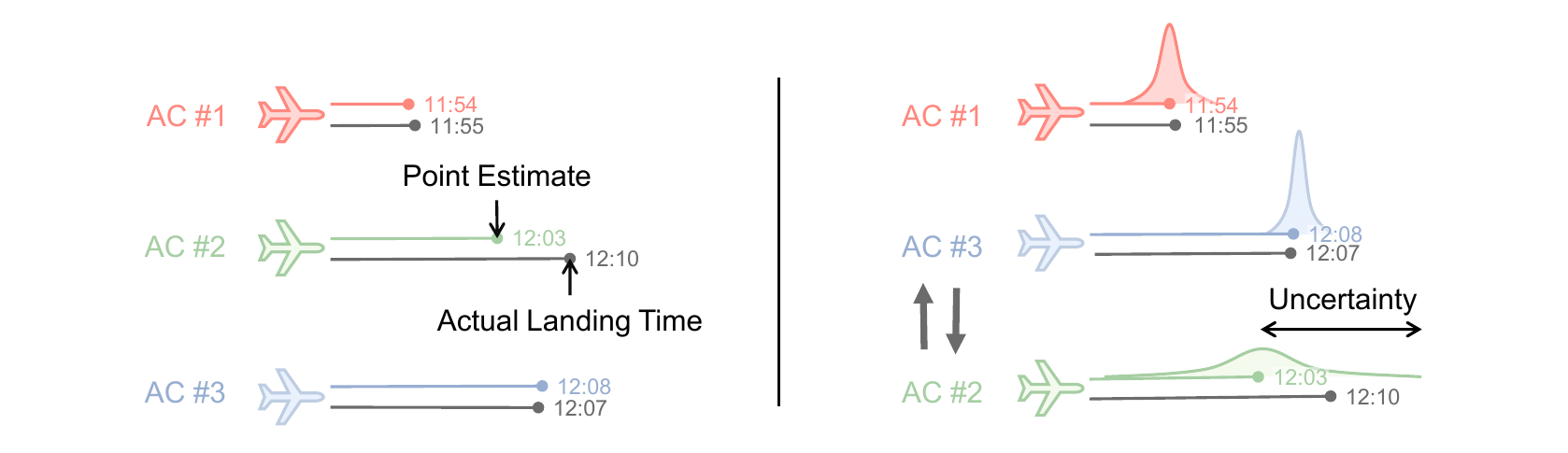}
    \vspace{-25pt}
    \caption{Comparison of deterministic (left) and probabilistic (right) predictions in estimating the arrival sequence.}
    \label{motive}
\end{figure} 
\vspace{5pt}

As illustrated in Figure~\ref{motive}, scheduling AC2 before AC3 based solely on a point estimate of their landing times can result in unnecessary delays and inefficiencies if the actual arrival order differs from the predictions. By contrast, when uncertainty estimates are available, airport decision-makers can interpret predictions more effectively. Although the point estimates suggest that AC2 will arrive before AC3, the different uncertainty levels associated with each aircraft may lead decision makers to anticipate that AC3 could arrive earlier than AC2.

In this paper, we propose a probabilistic multi-agent aircraft landing time prediction model that outputs a Gaussian distribution of the landing time for each aircraft, providing both the mean and standard deviation. The proposed model is based on our previous study on a Multi-Agent Inverted Transformer (MAIFormer) architecture~\cite{yoon2025maiformer}, which jointly learns individual aircraft trajectories and multi-agent aircraft interactions. By design, the model also produces agent-to-agent attention scores that can be used to interpret both its behavior and prediction outcomes.

The remainder of this paper is organized as follows: Section~\ref{metho} introduces the proposed probabilistic multi-agent aircraft landing time prediction model. Section~\ref{setup} describes the dataset, implementation details, and evaluation metrics. Section~\ref{result} presents the experimental results and analyses. Finally, Section~\ref{conclusion} concludes the paper with a summary and discussion of potential future research directions.

\section{Methodology}\label{metho}
\subsection{Problem Statements}
Multi-agent aircraft landing time prediction aims to jointly predict the time-to-land for multiple aircraft based on their past trajectories. Let $X = (x_1, x_2, \ldots, x_N) \in \mathbb{R}^{N\times T\times 3}$ be the observed air traffic scene involving $N$ aircraft flying over $T$ timestamps where $x_n \in \mathbb{R}^{T\times 3}$ represents the three-dimensional (3D) trajectory (latitude, longitude, and altitude) of the $n$-th aircraft. The prediction target is $Y = (y_1, y_2, \ldots, y_N) \in \mathbb{R}^{N}$, where each $y_n \in \mathbb{R}$ corresponds to the remaining flight time (i.e., the time until landing) for the $n$-th aircraft at the final observation timestamp $T$ (i.e., the current time). The landing time predictions are then given by the current time plus the predicted remaining flight time $\hat{Y} = (\hat{y}_1, \hat{y}_2, \ldots, \hat{y}_N) \in \mathbb{R}^{N}$. 

We adopt a probabilistic framework that models the distributions of landing times for multiple aircraft to incorporate the uncertainties in aircraft landing times. The objective is, then, to train a model with learnable parameters $\theta$ that provides the conditional distribution $P_{\theta}(Y | X)$. Based on $P_{\theta}$, we can predict the most likely aircraft landing times and quantify the uncertainty associated with those predictions. Our overall learning problem is defined as:
\begin{equation}
\theta^* = \text{arg}\min_{\theta}  \mathcal{L}(\theta) = -\log P_{\theta}(Y | X)
\end{equation}
The learned distribution $P_\theta$ maximizes the likelihood of the true landing times while providing calibrated uncertainty estimates by minimizing the negative log-likelihood (NLL).

\subsection{Proposed Framework}
\begin{figure}[t!]
    \centering
    \includegraphics[width=1.0\textwidth]{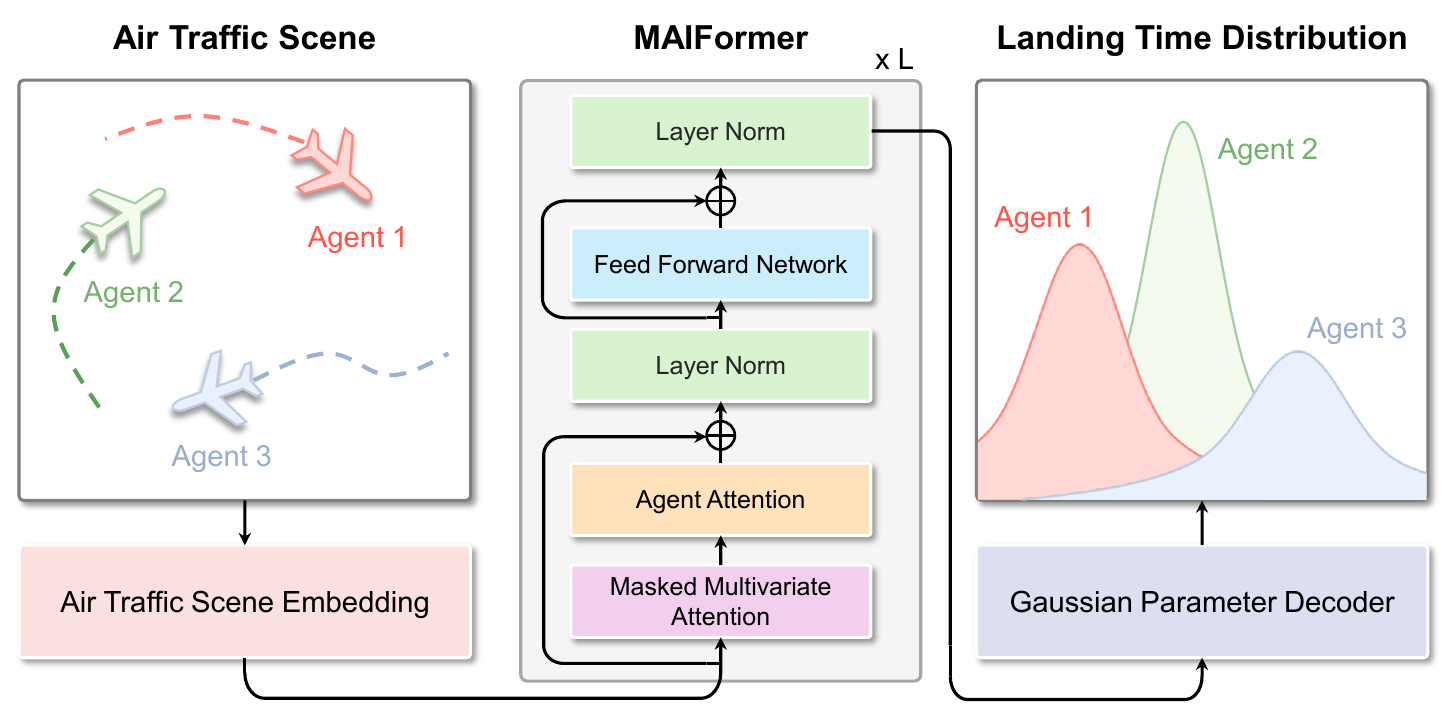}
    \caption{Overview of the proposed probabilistic multi-agent aircraft landing time prediction framework.}
    \vspace{-15pt}
    \label{archi}
\end{figure}

In this section, we present the proposed framework, which comprises three main components: (i) air traffic scene embedding, (ii) multi-agent trajectory encoder, and (iii) Gaussian parameter decoder (GPD). An overview of the framework for probabilistic multi-agent aircraft landing time prediction is shown in Figure~\ref{archi}, and we describe each component in detail below.

\vspace{-15pt}
\subsubsection{Air Traffic Scene Embedding}
We first reshape and invert $X$ from $\mathbb{R}^{N \times T \times 3}$ to $\mathbb{R}^{(N \times 3) \times T}$ to tokenize the trajectories of $N$ agents in the observed air traffic scene $X$. The resulting inverted air traffic scene $\bar{X}$ can then be considered as a single multivariate time series sequence with $N \times 3$ variates and a length of $T$~\cite{liu2023itransformer,yoon2025maiformer}. We then embed $\bar{X}$ into the latent dimension $D$ along the time axis using a linear layer:
\begin{equation}
    C = \text{SceneEmbedding}(\bar{X})
\end{equation}
where $C=(c^1_1,c^2_1,c^3_1,\ldots,c^1_N,c^2_N,c^3_N) \in \mathbb{R}^{(N\times3)\times D}$ is the embedded sequence of variate tokens, where each aircraft is represented by three variate tokens, the dimensions of which are $D$.

\begin{figure}[t!]
    \centering
    \includegraphics[width=1.0\textwidth]{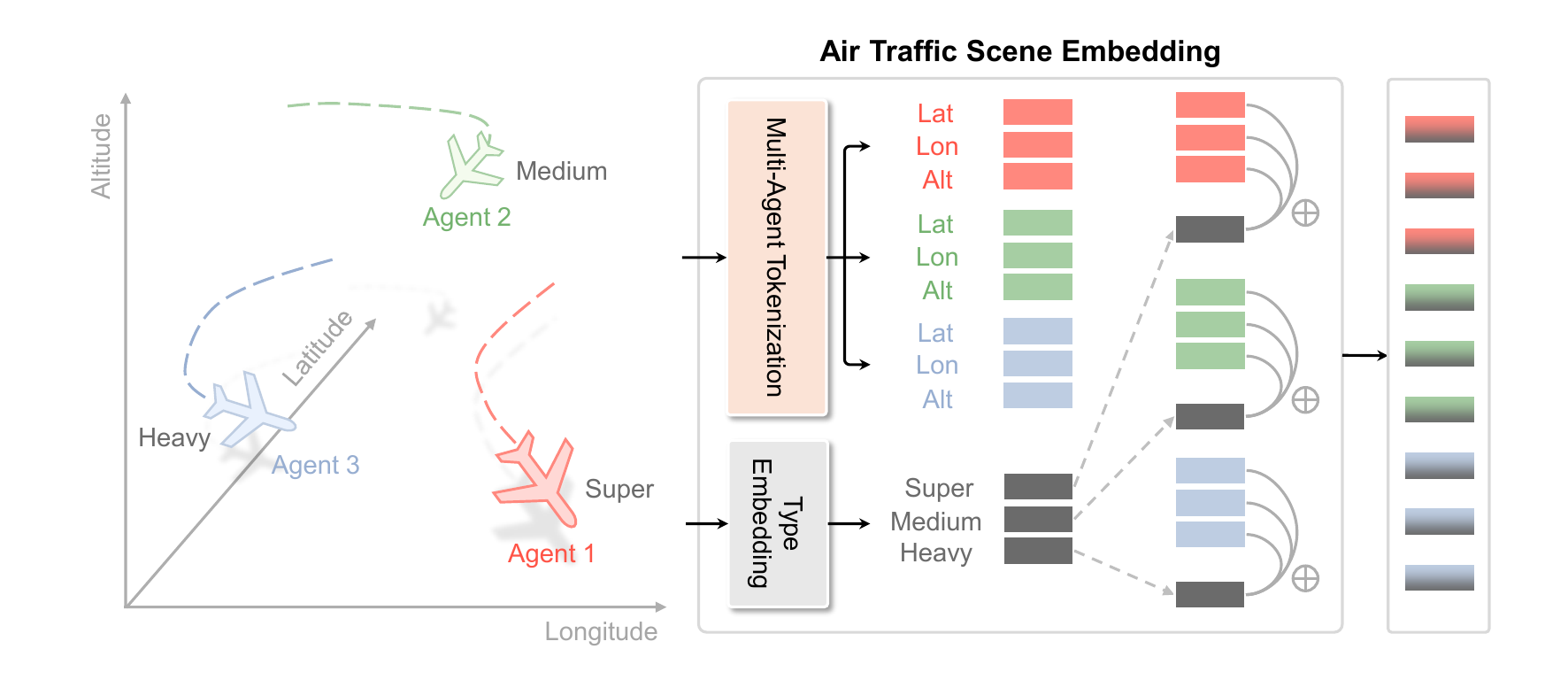}
    \vspace{-20pt}
    \caption{Illustration of air traffic scene embedding. For illustration, the number of aircraft is set to three, with each aircraft represented by three variates (latitude, longitude, and altitude).}
    \vspace{-15pt}
    \label{embedding}
\end{figure}

Next, we apply type embedding, which encodes each aircraft’s type (light, medium, heavy, or super) according to their wake turbulence category (WTC). We embed each aircraft’s type into a $D$-dimensional vector and, then, add it to $C$ in element-wise to incorporate the WTC information. Because the required time separations between consecutive arrivals depend on the WTCs of the aircraft, type embedding allows the model to learn the type-aware dynamics between aircraft. Figure~\ref{embedding} illustrates the process of air traffic scene embedding. 

\vspace{-15pt}
\subsubsection{Multi-Agent Trajectory Encoder}
Our multi-agent trajectory encoder is based on our previous work on MAIFormer~\cite{yoon2025maiformer}, which is built with $L$ stacked layers, each comprising two core modules: (i) masked multivariate attention and (ii) agent attention. The masked multivariate attention aims to model the movement of each individual aircraft, whereas the agent attention aims to capture the social interactions between them. The foundation of both the modules lies in the self-attention mechanism of Transformers~\cite{vaswani2017attention}.

\textbf{Self-Attention.} Self-attention is a mechanism that allows each element of a sequence to attend to all other elements in the sequence, capturing their contextual dependencies. Given an input sequence $X$, it is projected into three different vectors: queries, keys, and values, which are defined as $Q=XW^{Q},K= XW^{K}$, and $V=XW^{V}$, respectively, where $W$ are learnable weight matrices. Then, the self-attention computes the similarity between the queries and keys using dot products. These attention scores are scaled by $\sqrt{d_k}$, where $\sqrt{d_k}$ is the dimensionality of the keys, and then normalized using the softmax function. The normalized scores are subsequently used to compute a weighted sum of the values $V$ to aggregate the information from all the elements in the sequence. This process can be expressed as follows:
\begin{equation}
\text{Attention}(Q,K,V) = \text{Softmax}\left(\frac{QK^{\top}}{\sqrt{d_k}}\right)V
\label{eq:Attention}
\end{equation}
This results in context-rich representations, where each element selectively integrates information from the entire sequence based on learned attention weights.

\vspace{5pt}
\begin{figure}[h!]
    \centering
    
    \begin{subfigure}[t]{0.48\textwidth}
        \centering
        \includegraphics[width=\textwidth]{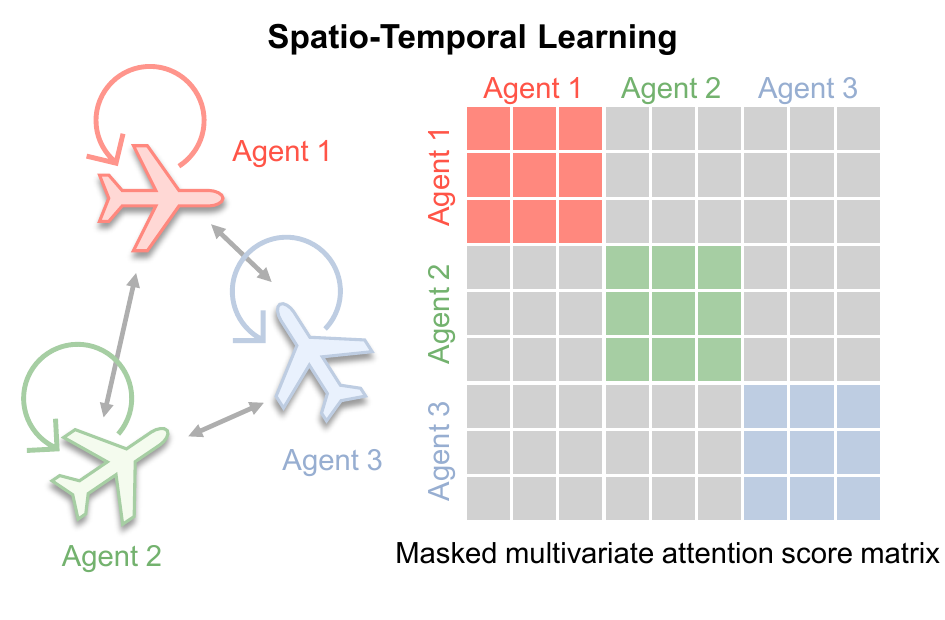}
        \vspace{-20pt}
        \caption{Masked multivariate attention}
        \label{MMA}
    \end{subfigure}
    \hfill
    \begin{subfigure}[t]{0.48\textwidth}
        \centering
        \includegraphics[width=\textwidth]{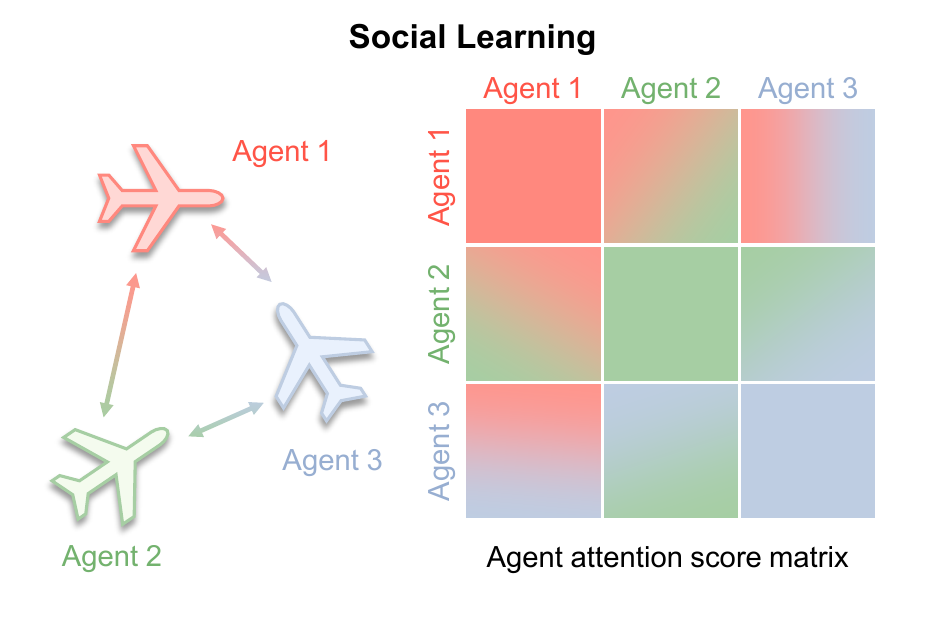}
        \vspace{-20pt}
        \caption{Agent attention}
        \label{AA}
    \end{subfigure}
    \vspace{-5pt}
    \caption{Illustration of masked multivariate attention and agent attention. The colored arrows denote where attention is applied while the gray arrows indicate where attention is masked.}
    \vspace{-15pt}
    \label{fig:MMA_AA}
\end{figure}

\textbf{Masked Multivariate Attention.} Building on the concept of self-attention, this module is designed to learn the spatio-temporal (ST) patterns of each aircraft trajectory. Specifically, in masked multivariate attention, the self-attention is applied only within the variates (latitude, longitude, and altitude) of the $i$-th aircraft, $C_i=(c^1_i,c^2_i,c^3_i) \in \mathbb{R}^{3\times D}$. This design ensures that the module learns multivariate correlations within an individual aircraft without interference from others. To mask out cross-aircraft interactions, the $i$-th aircraft's variates are restricted from attending to those of the $j$-th aircraft ($i \neq j$), as illustrated in Figure~\ref{MMA}. Formally, the masked multivariate attention is defined as:
\begin{align}
    &\text{MaskedMultivariateAttention}(Q_{\text{ST}},K_{\text{ST}},V_{\text{ST}}) 
    = \text{Softmax}\!\left(\frac{Q_{\text{ST}}K_{\text{ST}}^{\top} \odot M}{\sqrt{d_k}}\right)V_{\text{ST}} \label{eq:mma1} \\[3pt]
    &Q_{\text{ST}},K_{\text{ST}},V_{\text{ST}} 
    = CW^Q_{\text{ST}}, \;CW^K_{\text{ST}}, \; CW^V_{\text{ST}} \label{eq:mma2} \\[3pt]
    &M[m, n]  
    = 
    \begin{cases}
        1, & \text{if } \left\lfloor \tfrac{m}{3} \right\rfloor = \left\lfloor \tfrac{n}{3} \right\rfloor \\[6pt]
        -\infty, & \text{otherwise}, \quad \forall \, m, n \in \{0,1,\dots,(N \times 3)-1\}
    \end{cases} \label{eq:mma3}
\end{align}
where $M$ is a mask matrix in which the cross-aircraft elements are set to be $-\infty$. During the softmax operation, these masked elements become zero, ensuring that each aircraft attends only to its own variates. The resulting representation is  $C_{\text{ST}}=(c^1_{{\text{ST},1}},c^2_{{\text{ST},1}},c^3_{{\text{ST},1}},\ldots,c^1_{{\text{ST},N}},c^2_{{\text{ST},N}},c^3_{{\text{ST},N}}) \in \mathbb{R}^{(N\times3)\times D}$.

\textbf{Agent Attention.} After capturing individual aircraft behavior through the masked multivariate attention module, the agent attention module models the social interaction (SC) patterns among aircraft. This module allows each agent to attend to all other agents, enabling MAIFormer to learn the influence of the aircraft on each other. Figure~\ref{AA} illustrates the agent attention process. As the information of each aircraft is represented by three variate tokens, we concatenate the three $D$-dimensional tokens $C_{\text{ST},i}=(c_{\text{ST},i}^1,c_{\text{ST},i}^2,c_{\text{ST},i}^3) \in \mathbb{R}^{3\times D}$ into a single agent token of dimension $C_{\text{A},i}=(c_{\text{ST},i}^1 \oplus c_{\text{ST},i}^2 \oplus c_{\text{ST},i}^3) \in \mathbb{R}^{1\times (3\times D)}, \ i=1,2,\ldots,N$. This aggregation provides direct aircraft-to-aircraft attention scores that enhance the interpretability of the model. After collecting all agent tokens into the agent sequence $C_\text{A} =(C_{\text{A},1},C_{\text{A},2},\ldots,C_{\text{A},N})\in \mathbb{R}^{N\times(3\times D)}$, the agent attention is applied as follows:
\begin{align}
    &\text{AgentAttention}(Q_{\text{SC}},K_{\text{SC}},V_{\text{SC}}) = \text{Softmax}\!\left(\frac{Q_{\text{SC}}K_{\text{SC}}^{\top}}{\sqrt{d_k}}\right)V_{\text{SC}} \label{eq:aa1} \\[3pt]
    &Q_{\text{SC}},K_{\text{SC}},V_{\text{SC}} 
    = C_AW^Q_{\text{SC}}, \;C_AW^K_{\text{SC}}, \; C_AW^V_{\text{SC}} \label{eq:aa2} 
\end{align}
The output of the agent attention module is $C_{\text{SC}} =(C_{\text{SC},1},C_{\text{SC},2},\ldots,C_{\text{SC},N})\in \mathbb{R}^{N\times(3\times D)}$, which is reshaped back to $\mathbb{R}^{(N\times3)\times D}$ after the subsequent operations of layer normalization and FFN.

\vspace{-15pt}
\subsubsection{Gaussian Parameter Decoder}
\begin{figure}[t!]
    \centering
    \includegraphics[width=1.0\textwidth]{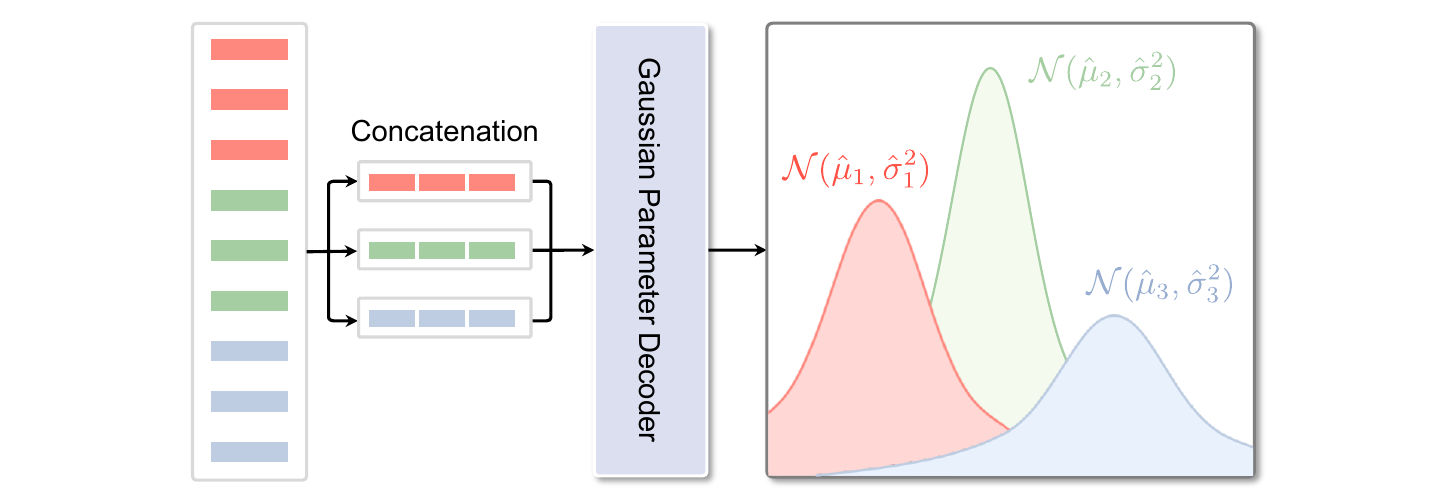}
    \vspace{-15pt}
    \caption{Illustration of Gaussian parameter decoder.}
    \vspace{-15pt}
    \label{GPD}
\end{figure}

After the air traffic scene $X$ is encoded through the stacked $L$ multi-agent encoder layers, the GPD is applied to output the predicted mean $\hat{\mu}_i$ and standard deviation $\hat{\sigma}_i$ of landing time for each aircraft. Rather than modeling a full joint probability distribution across all agents, the GPD produces a marginal probabilistic prediction for the landing time $y_i$ of each agent, parameterized by the predicted mean $\hat{\mu}_i$, and standard deviation $\hat{\sigma}_i$:
\begin{equation}
    p(y_i)= \mathcal{N}(y_i \vert \hat{\mu}_i, \hat{\sigma}^2_i) , \quad i \in 1, \ldots, N.
\end{equation}
In this formulation, the GPD assumes conditional independence among the $y_i$. Nevertheless, the landing time prediction for each aircraft remains contextually dependent on the shared multi-agent representations produced by the previous multi-agent encoder. Although full covariance modeling can better capture explicit inter-agent uncertainty correlations, it introduces significant computational complexity. Figure~\ref{GPD} presents the overview of the GPD process.

\vspace{-15pt}
\subsubsection{Training Objective}
To train the model with probabilistic outputs, we adopt NLL as the loss function to maximize the likelihood of the actual landing time $y_i$ given the predicted mean $\hat{\mu}_i$ and standard deviation $\hat{\sigma}_i$. The NLL is defined as: 
\begin{equation}
\mathcal{L}_{\text{NLL}}
= -\frac{1}{N}\sum_{i=1}^{N} \log p(y_i \mid \hat{\mu}_i, \hat{\sigma}_i^2)
= \frac{1}{N} \sum_{i=1}^{N}
\underbrace{
  \frac{1}{2} \log (2\pi\hat{\sigma}_i^2) 
  \vphantom{\frac{(y_i - \hat{\mu}_i)^2}{2\hat{\sigma}_i^2}}
}_{\text{uncertainty penalty}}
+
\underbrace{
  \frac{(y_i - \hat{\mu}_i)^2}{2\hat{\sigma}_i^2}
}_{\text{prediction error}}
\end{equation}
The NLL loss can be interpreted as the sum of the two components. The first term, labeled as uncertainty penalty, penalizes the model for predicting excessively large variances $\hat{\sigma}_i^2$. The second term, denoted as prediction error, corresponds to the squared deviation between the true landing time $y_i$ and the predicted mean $\hat{\mu}_i$, normalized by the predicted variance. This term enforces the accuracy of the mean prediction while encouraging proper calibration of the predicted variance. Together, these two components balance the trade-off between accuracy and uncertainty, ensuring that the model not only minimizes the prediction error but also provides reliable estimates of confidence in its output.

Furthermore, we normalize the total NLL by averaging over all $N$ aircraft rather than summing them. This normalization ensures that the scale of the loss remains consistent and comparable across air traffic scenes with different numbers of aircraft, thereby stabilizing the training process.

\section{Experimental Setup}\label{setup}
\subsection{Dataset Description and Pre-processing}
In this work, we used an ADS-B trajectory dataset collected from the terminal airspace of Incheon International Airport (ICN) in South Korea, obtained via FlightRadar24\footnote{\textcolor{DodgerBlue}{\url{https://www.flightradar24.com}}}, covering January to December 2023. The ADS-B trajectory data included information such as aircraft type, a sequence of 3D positions, and departure and arrival airports. For landing time prediction, we extracted all arrival trajectories to the ICN. While not entirely independent, arrivals and departures are spatially segregated and managed by different ATCo, which makes it reasonable to focus solely on arrivals with minimal loss of generality. The trajectories were truncated to begin at the 70-nautical-mile boundary from the airport reference point of the ICN and resampled at a uniform 6-second interval using a piecewise cubic Hermite interpolating polynomial (PCHIP)~\cite{fritsch1980monotone} to address the irregular sampling rates in the ADS-B dataset. 

The dataset was then chronologically split into training, validation, and test sets at an 8:1:1 ratio. We applied a sliding window approach with a stride of 6 seconds to construct the multi-agent dataset, where each window spanned 2 minutes of trajectories (20 time steps) over the traffic scene. Each resulting scene contains the multi-agent trajectories along with their corresponding remaining flight times.

\subsection{Implementation Details}
We set the number of multi-agent encoder layers $L$ to three. The latent dimension of the model $D$ was set to 256, and both the masked multivariate attention and agent attention modules employed 8 attention heads. 
The inner-layer dimension of the FFN in the encoder was set to 1024, with Gaussian error linear units (GELUs)~\cite{hendrycks2016gaussian} to introduce nonlinearity. The GPD has four linear layers with GELUs, progressively reducing the dimensionality from 768 ($256 \times 3$) to 512, 256, 128, and finally to 2, which correspond to the predicted mean and standard deviation values.

We trained the model for 300 epochs with a batch size of 64 air traffic scenes using the AdamW optimizer~\cite{loshchilov2017adamw}. The initial learning rate was set to $10^{-4}$ and decayed to $10^{-6}$ within each 100-epoch cycle, after which it was reset to $10^{-4}$ at the beginning of subsequent cycles. We also applied a dropout rate of 0.1 throughout the model for regularization. The experiments were implemented in a software environment of Python 3.10.18 and PyTorch 2.8.0~\cite{paszke2019pytorch} with CUDA 12.8, running on a desktop equipped with an Intel Core i7 processor, 32 GB of RAM, and an NVIDIA RTX 4060 GPU.

\subsection{Comparison Baselines}
We compared the performance of the proposed landing time prediction model with that of several baselines commonly referenced in previous works. First, we included a multiple linear regression (MLR) model as the standard approach for aircraft landing time prediction~\cite{hong2015trajectory}. Second, we considered a gradient boosting machine (LightGBM)~\cite{NIPS2017_6449f44a}, which has also demonstrated strong performance in this task~\cite{wang2020automated}. Finally, we evaluated extreme gradient boosting (XGBoost)~\cite{chen2016xgboost}, a nonlinear ensemble model that enhances LightGBM with improved regularization and computational efficiency.

To assess the landing time prediction accuracy, we used the mean absolute error (MAE), root mean squared error (RMSE), and mean absolute percentage error (MAPE). MAE provides the average absolute prediction error and serves as a general performance indicator, whereas RMSE squares the prediction errors before averaging and is more sensitive to large errors. MAPE measures the prediction error as a percentage of the actual value. These metrics are defined as follows:
\begin{align}
    &\text{MAE}  = \frac{1}{S} \sum_{i=1}^{S} \left| y_i - \hat{y}_i \right| \\
    &\text{RMSE} = \sqrt{ \frac{1}{S} \sum_{i=1}^{S} \left(  y_i - \hat{y}_i  \right)^2 } \\
    &\text{MAPE} = 100\% \times \frac{1}{S} \sum_{i=1}^{S} \left| \frac{ y_i - \hat{y}_i }{y_i} \right|
\end{align}
where $S$ denotes the number of flight trajectories in the test dataset, and $y_i$ and $\hat{y}_i$ denote the actual and predicted remaining flight times, respectively.

In addition, the arrival sequence induced by the predicted landing times $\widehat{Seq}$ was compared with the actual arrival sequence $Seq$. This evaluation is important because, even with small prediction errors, the model can produce incorrect arrival sequences in a complex airspace where multiple aircraft enter from multiple entry fixes with only minimal separations. To quantify the consistency between $\widehat{Seq}$ and $Seq$, we used Spearman’s rank correlation coefficient~\cite{hullermeier2008label} and Kendall rank correlation coefficient~\cite{kendall1938new}, following previous studies~\cite{jung2018data,choi2023multi}. Spearman’s rank correlation coefficient $\rho$ evaluates the degree of similarity between two rankings by measuring the differences of their rank values, whereas Kendall rank correlation coefficient $\tau$ quantifies the consistency of the pairwise ordering between them. The two metrics are computed as follows:
\begin{equation}
    \rho = 1 - \frac{6D(Seq,\widehat{Seq})}{N(N^2 - 1)}
\end{equation}
\begin{equation}
    \tau = \frac{N_c - N_d}{\binom{N}{2}}
\end{equation}
where $D(Seq,\widehat{Seq}) \equiv \sum_{i=1}^{N} \big( Seq(AC_i) - \widehat{Seq}(AC_i) \big)^2$ denotes the rank distance between the actual and predicted sequences, $N$ denotes the number of aircraft, $N_c$ denotes the number of concordant pairs, and $N_d$ is the number of discordant pairs. Both $\rho$ and $\tau$ range from $-1$ to $+1$, where $-1$ denotes that $\widehat{Seq}$ is the exact reverse of $Seq$, and $+1$ denotes that they are identical.

\section{Results}\label{result}
\subsection{Performance Analysis}\label{quant}
Table~\ref{tab:eta_performance1} summarizes the prediction performances of our model and the baselines. We also reported the performance improvement (PI) as the relative gain over the second-best baseline, defined as $|S_2 - S_1| / S_2$ for error-based metrics (MAE, RMSE, MAPE; lower is better) and $|S_1 - S_2| / S_2$ for rank-based metrics (Spearman's $\rho$, Kendall's $\tau$; higher is better), where $S_1$ denotes the score of the proposed model and $S_2$ denotes the score of the second-best baseline. Note that in the following analyses, the predicted mean $\hat{\mu}_i$ is used as the maximum likely predicted value for the remaining flight time $\hat{y}_i$.

\vspace{5pt}
\begin{table}[h]
\caption{Landing time prediction performance and arrival sequence consistency.}
\vspace{-5pt}
\centering
%\small
\resizebox{0.8\textwidth}{!}{%
\begin{tabular}{c|c|c|c|c|c}
\toprule
Model & MAE (sec) $\downarrow$ & RMSE (sec) $\downarrow$ & MAPE (\%) $\downarrow$ & Spearman ($\rho$) $\uparrow$ & Kendall ($\tau$) $\uparrow$\\
\midrule
MLR & 113.1209 & 150.8267 & 63.77 & 0.926 & 0.875 \\
LightGBM & 57.1494 & 87.5603 & 12.88 & 0.988 & 0.975\\
XGBoost & 55.7939 & 81.9525 & 16.38 & 0.989 & 0.978 \\
\cellcolor{limegreen!20}\textbf{Ours} & \cellcolor{limegreen!20}\textbf{38.1099} 
     & \cellcolor{limegreen!20}\textbf{58.5113} 
     & \cellcolor{limegreen!20}\textbf{6.22} 
     & \cellcolor{limegreen!20}\textbf{0.991} 
     & \cellcolor{limegreen!20}\textbf{0.982}\\
\midrule
PI (\%) & 31.69 & 28.60 & 51.70 & 0.20 & 0.40 \\
\bottomrule
\end{tabular}
}
\vspace{2pt}
\label{tab:eta_performance1}
\end{table}

As summarized in the table, the proposed model outperformed all the baseline models. In terms of the prediction accuracy, the MAE, RMSE, and MAPE of the proposed model achieved PI values of 31.69\%, 28.60\%, and 51.70\%, respectively. Specifically, MLR exhibited the lowest performance due to its assumption of a purely linear relationship between aircraft positions and remaining flight time. By contrast, LightGBM and XGBoost achieved better performance by capturing more complex dependencies. However, because MLR, LightGBM, and XGBoost cannot account for multi-agent interactions, their predictive capabilities remain limited. In comparison, our multi-agent landing time prediction model achieved the best results, as it explicitly accounted for inter-agent interactions within the air traffic scene and learned the underlying air traffic control patterns from historical data. The MAE of 38.1099 seconds demonstrates that our model provides highly accurate landing times, whereas the low RMSE and MAPE values further confirm its robustness and stability across diverse traffic scenarios.

\begin{figure}[t!]
    \centering
    
    \begin{subfigure}[t]{0.48\textwidth}
        \centering
        \includegraphics[width=\textwidth]{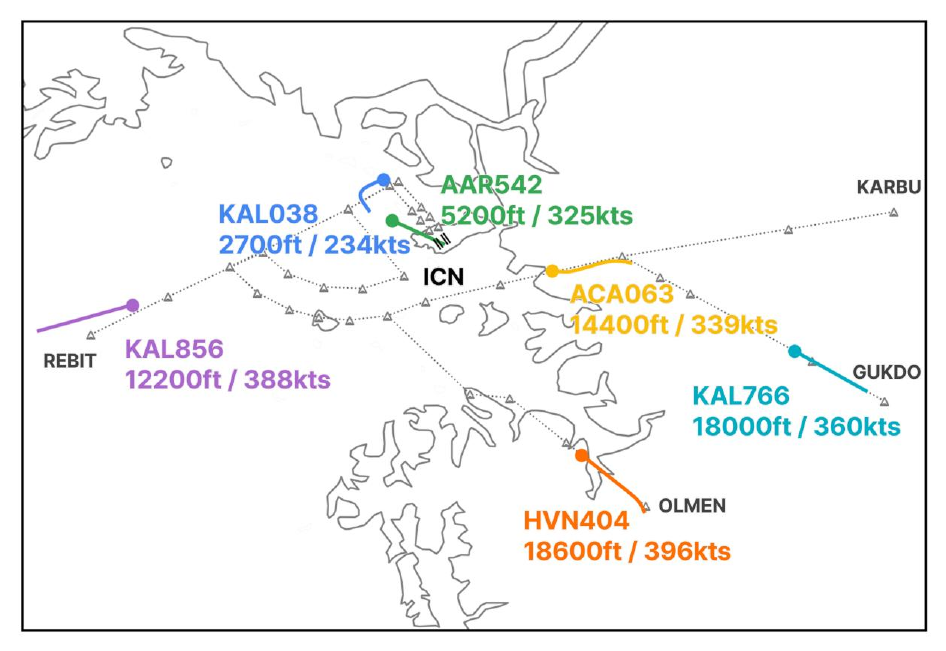}
        \caption{Example air traffic scenario}
        \label{traffic_scene}
    \end{subfigure}
    \hfill
    \begin{subfigure}[t]{0.48\textwidth}
        \centering
        \raisebox{-0.045cm}{\includegraphics[width=\textwidth]{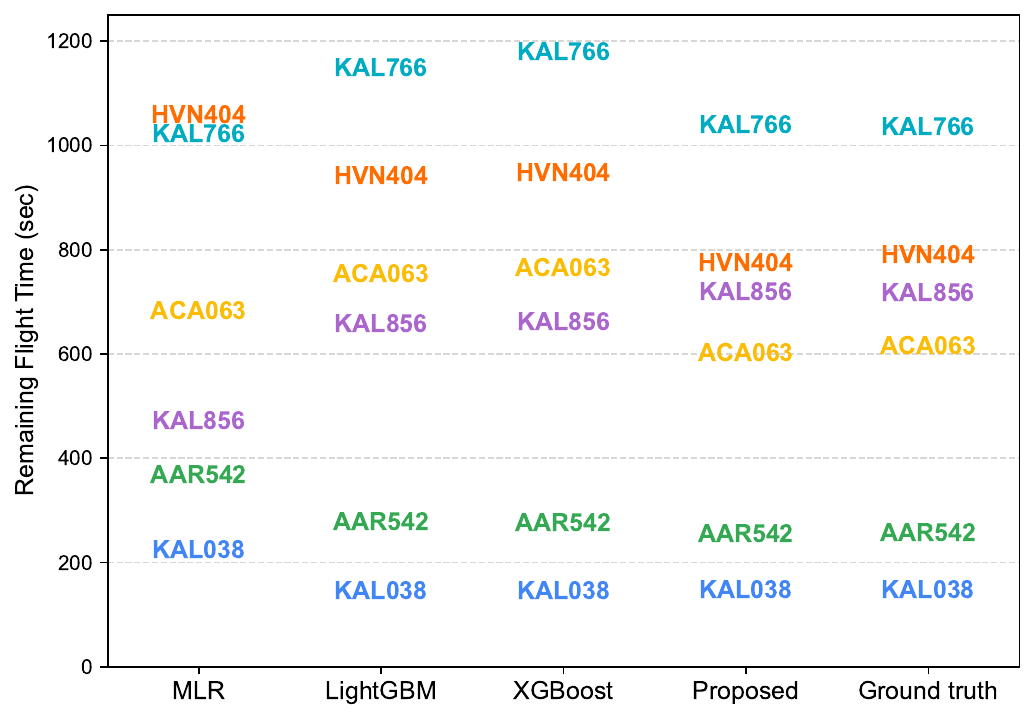}}
        \caption{Induced arrival sequence based on prediction results}
        \label{seq}
    \end{subfigure}
    \vspace{-5pt}
    \caption{Illustrative example of arrival sequence consistency analysis.}
    \label{fig:sequence}
    \vspace{-15pt}
\end{figure}
\vspace{5pt}

Regarding arrival sequence consistency, our model achieved the highest scores for both Spearman's and Kendall's correlations, as summarized in Table~\ref{tab:eta_performance1}. Although the other models demonstrated reasonable consistency with the actual sequences, they frequently suffered from incorrect arrival sequence predictions. Figure~\ref{traffic_scene} illustrates an example traffic scenario in which multiple aircraft enter the terminal airspace of ICN through different entry fixes (REBIT, OLMEN, GUKDO, and KARBU). Each aircraft’s current position is marked with a circle, and its past 2-minute trajectory is depicted as a tail. As shown in Figure~\ref{seq}, the landing time predictions generated by MLR, LightGBM, and XGBoost were generally accurate, resulting in mostly consistent arrival sequences; however, the actual and predicted sequences differed slightly between \textcolor[HTML]{AA66CC}{\texttt{KAL856}} and \textcolor[HTML]{FBBC05}{\texttt{ACA063}}. Considering the preceding traffic flow, ATCo issued a shortcut to \textcolor[HTML]{FBBC05}{\texttt{ACA063}}, because sufficient spacing existed with its preceding aircraft, \textcolor[HTML]{34A853}{\texttt{AAR542}}. However, because the single-agent models (MLR, LightGBM, and XGBoost) cannot capture such interactions and because the distance along the route from \textcolor[HTML]{AA66CC}{\texttt{KAL856}} to ICN is shorter than that of \textcolor[HTML]{FBBC05}{\texttt{ACA063}}, they simply predict that \textcolor[HTML]{AA66CC}{\texttt{KAL856}} will land earlier. By contrast, our model correctly inferred that \textcolor[HTML]{FBBC05}{\texttt{ACA063}} would land first, followed by \textcolor[HTML]{AA66CC}{\texttt{KAL856}}, by leveraging ATCo behavior patterns such as radar vectoring and sequencing learned from historical data in similar scenarios.

\subsection{Qualitative Analysis}\label{quali}
\begin{figure}[h!]
    \centering
    \begin{subfigure}[t]{0.48\textwidth}
        \centering
        \includegraphics[width=\textwidth]{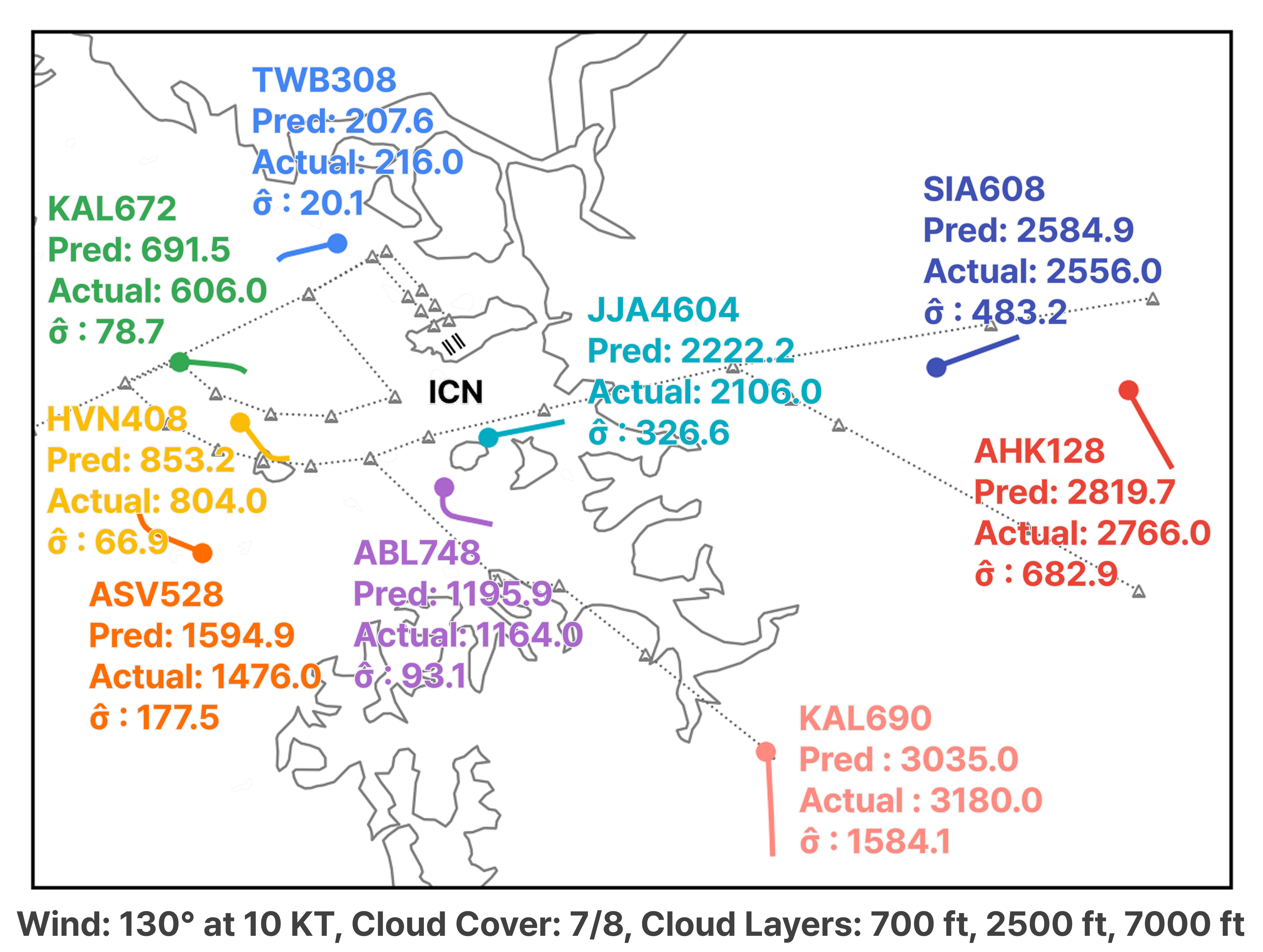}
        \caption{Unstructured air traffic scenario}
        \label{unstable_}
    \end{subfigure}
    \hfill
    \begin{subfigure}[t]{0.48\textwidth}
        \centering
        \includegraphics[width=\textwidth]{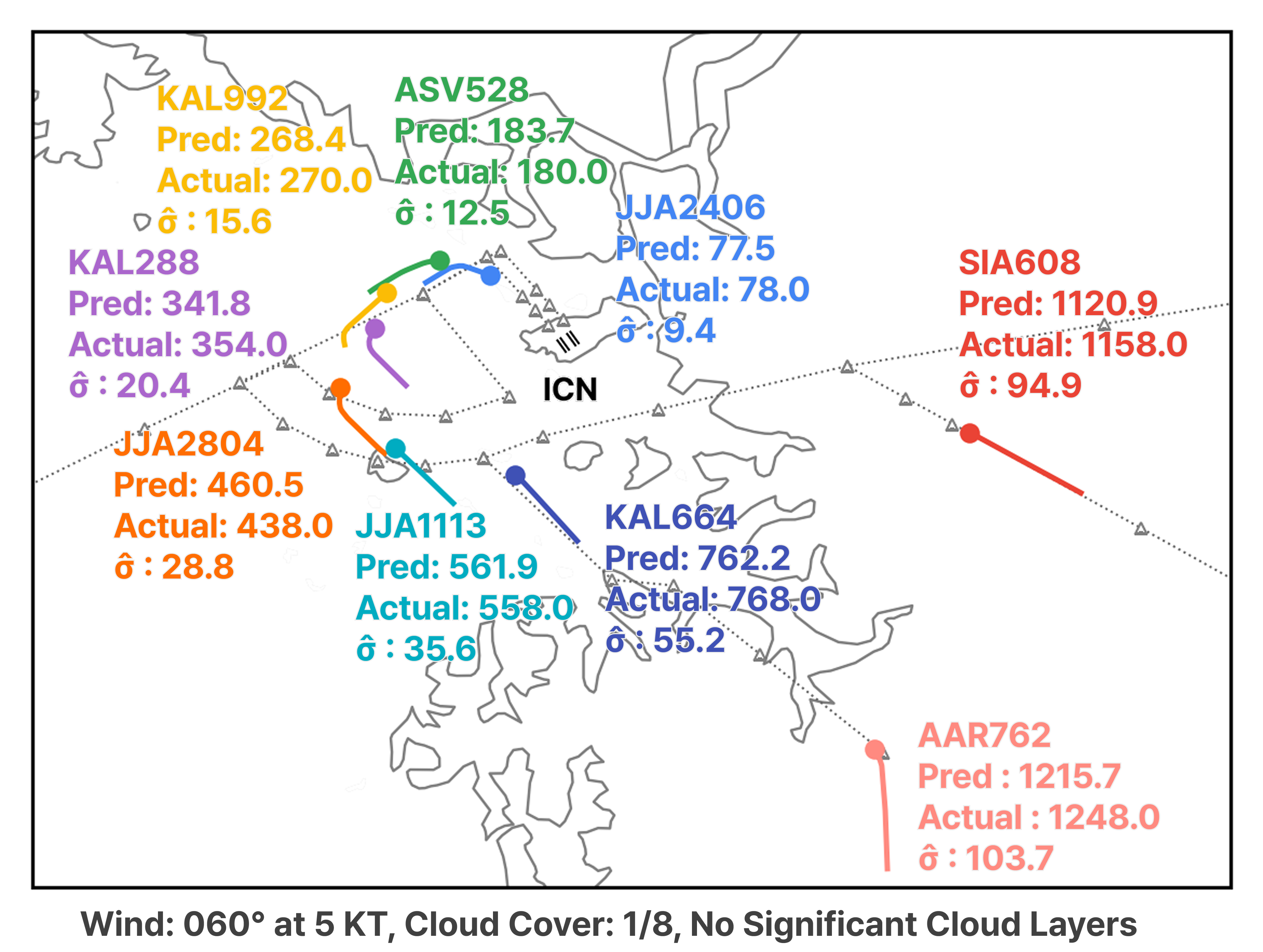}
        \caption{Structured air traffic scenario}
        \label{stable_}
    \end{subfigure}
    \vspace{-5pt}
    \caption{Illustrative examples of probabilistic landing time prediction.}
    \label{fig:unstable_stable}
\end{figure}

Figure~\ref{fig:unstable_stable} presents the multi-agent aircraft landing time prediction results for two traffic scenarios: the left figure depicts an unstructured traffic flow, whereas the right figure shows a structured traffic flow. A structured scenario is characterized by one or a few dominant traffic patterns, whereas an unstructured scenario involves multiple overlapping and irregular patterns. For each aircraft, its call sign, predicted and actual remaining flight times, and the associated prediction uncertainty generated by the proposed method are presented. Weather conditions including wind direction and speed, cloud cover, and cloud layers\footnote{Cloud cover refers to the fraction of the sky obscured by clouds, while cloud layers denote the altitude of distinct cloud strata.} are also shown below each figure, based on observation data from the aerodrome meteorological observing system (AMOS) at ICN.

As shown in Figure~\ref{unstable_}, \textcolor[HTML]{34A853}{\texttt{KAL672}} began deviating from its arrival route rather than following its preceding aircraft, \textcolor[HTML]{4285F4}{\texttt{TWB308}}, which was about to reach the initial approach fix and commence the landing procedure. We conjecture that adverse weather conditions (near-complete cloud cover, reported as 7/8, at 700 and 2500 ft) prompted the ATCo to impose a greater separation than usual between \textcolor[HTML]{4285F4}{\texttt{TWB308}} and \textcolor[HTML]{34A853}{\texttt{KAL672}} by vectoring \textcolor[HTML]{34A853}{\texttt{KAL672}} out of the cloud layers. This increased separation likely disrupted the flow and contributed to the emergence of more complex traffic patterns.

Consequently, the model predicted substantially delayed landing times for other arriving aircraft and introduced considerable uncertainty, particularly for those farther from the airport (\textcolor[HTML]{3F51B5}{\texttt{SIA608}}, \textcolor[HTML]{EA4335}{\texttt{AHK128}}, and \textcolor[HTML]{FF8A80}{\texttt{KAL690}}). These results are reasonable, as the deviation of \textcolor[HTML]{34A853}{\texttt{KAL672}} makes it inherently unclear when other aircraft will land and how the arrival sequence will be established. Nevertheless, despite the complexity of this situation, our model still generates highly accurate landing time predictions in terms of the most likely estimates, suggesting that it has effectively learned how ATCo handle such traffic situations from historical data.

In Figure~\ref{stable_}, the traffic scenario exhibits a more structured pattern, with most aircraft aligned on their arrival paths. The weather conditions were favorable (minimal cloud cover, reported as 1/8, no significant cloud layers), allowing the aircraft to maintain their nominal routes without disruption. Because this traffic pattern is highly structured, the model can accurately predict aircraft landing times, regardless of their distance from the airport, and the associated uncertainties remain extremely low. Notably, although \textcolor[HTML]{FF8A80}{\texttt{AAR762}} in Figure~\ref{stable_} is positioned similarly to \textcolor[HTML]{FF8A80}{\texttt{KAL690}} in Figure~\ref{unstable_}, the two cases produced markedly different prediction outcomes in terms of both mean and standard deviation. This contrast highlights the superiority of the proposed model in understanding different air traffic situations and demonstrates that its uncertainty estimates reflect the complexity and reality of actual operations.

\begin{figure}[t!]
    \centering
    \begin{subfigure}[t]{\textwidth}
        \centering
        \includegraphics[width=1.0\textwidth]{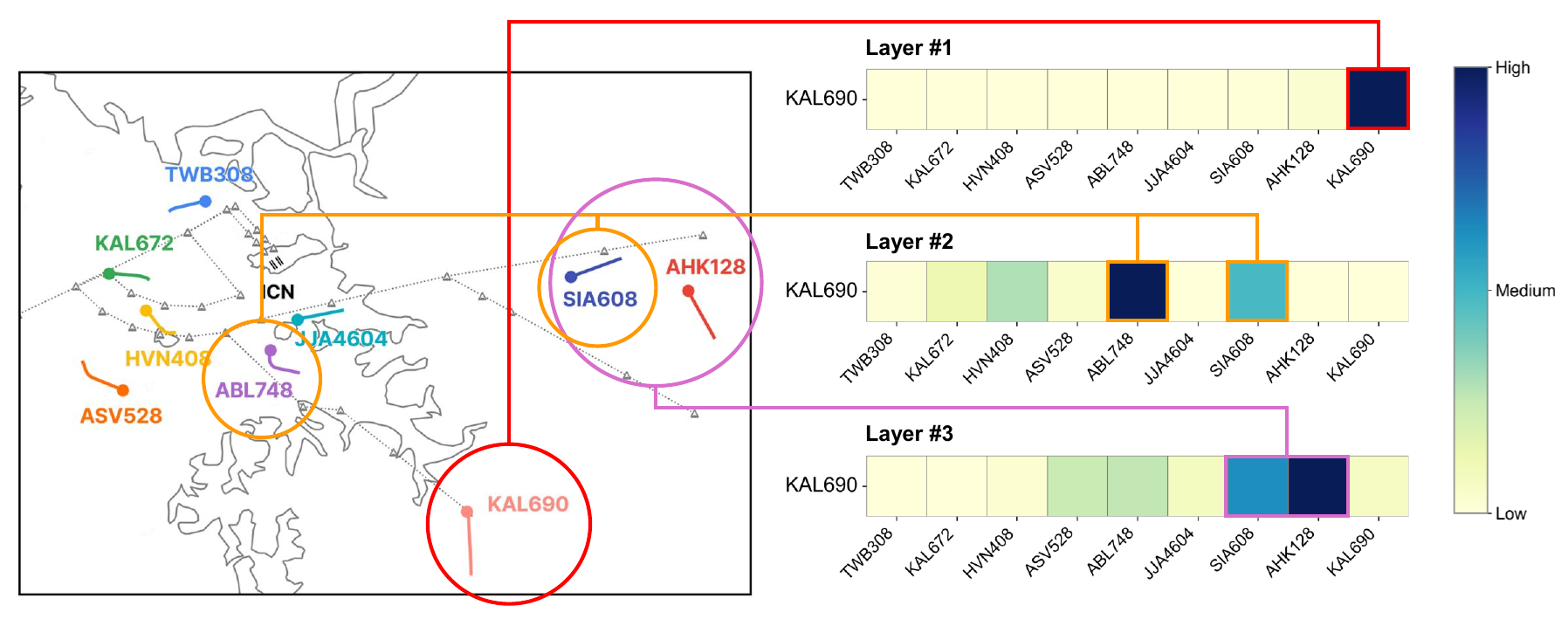}
        \caption{Unstructured air traffic scenario and agent attention scores}
        \label{unstable}
    \end{subfigure}
    
    \vspace{0.1cm} 

    \begin{subfigure}[t]{\textwidth}
        \centering
        \includegraphics[width=1.0\textwidth]{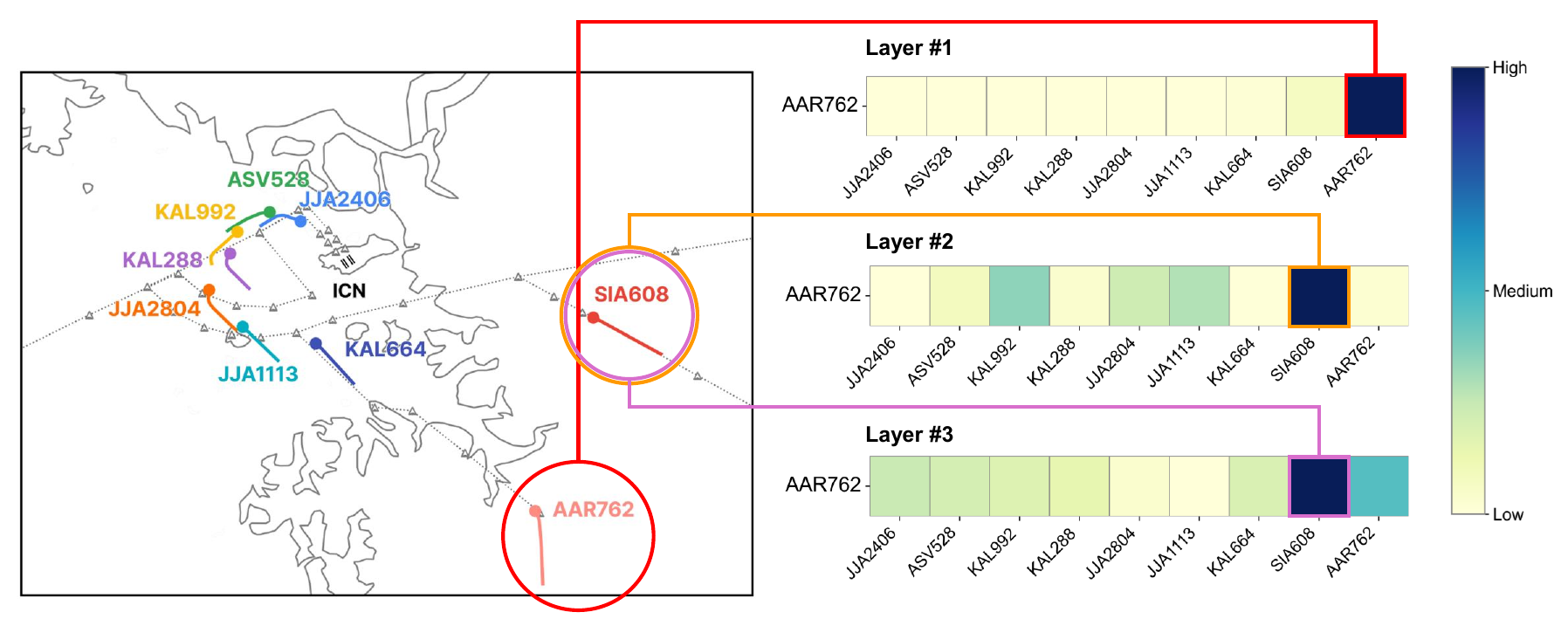}
        \caption{Structured air traffic scenario and agent attention scores}
        \label{stable}
    \end{subfigure}
    \vspace{-0.2cm}
    \caption{Layer-wise agent attention score analysis for two traffic scenarios in Figure~{\ref{fig:unstable_stable}}.}
    \vspace{-15pt}
    \label{fig:attn_compare}
\end{figure}

\subsection{Interpreting Model Behavior via Agent Attention Scores}\label{behave} 
Building on our previous works~\cite{yoon2025maiformer,chai2025learning}, the agent-to-agent attention scores produced by the proposed model can be used to explain air traffic scenes and interpret the model’s outcomes. As illustrative examples, we selected the two traffic scenarios shown in Figure~\ref{fig:unstable_stable}. The attention scores for these scenarios are presented in Figure~\ref{unstable} and~\ref{stable}; the left panel of each figure depicts the traffic scenario, and the right panel shows the agent attention scores across the three layers. For each layer, we examined how much attention the model assigns to other aircraft when predicting the landing time of the target aircraft.

In Figure~\ref{unstable}, we analyzed the attention scores from the perspective of \textcolor[HTML]{FF8A80}{\texttt{KAL690}} as the target aircraft. In the first layer, \textcolor[HTML]{FF8A80}{\texttt{KAL690}} primarily attends to itself. The model initially relies on its own movement and spatial location to form a preliminary representation for landing time prediction. In the subsequent layers, however, \textcolor[HTML]{FF8A80}{\texttt{KAL690}} began attending to \textcolor[HTML]{3F51B5}{\texttt{SIA608}} and \textcolor[HTML]{EA4335}{\texttt{AHK128}}, whose predicted landing times were similar to that of \textcolor[HTML]{FF8A80}{\texttt{KAL690}}. This suggests that the model then refines the preliminary representation by considering the interactions with nearby aircraft. Interestingly, \textcolor[HTML]{AB47BC}{\texttt{ABL748}} receives high attention in the second layer, despite not being adjacent to \textcolor[HTML]{FF8A80}{\texttt{KAL690}} in the estimated landing sequence. This may indicate that the model accounts for uncertainty arising from the vectoring of \textcolor[HTML]{AB47BC}{\texttt{ABL748}} for path stretching during an uncertain upstream arrival sequence, which can affect the trajectory of \textcolor[HTML]{FF8A80}{\texttt{KAL690}} and, consequently, its remaining flight time.

Figure~\ref{stable} shows a similar phenomenon when we analyzed the agent attention scores from the perspective of \textcolor[HTML]{FF8A80}{\texttt{AAR762}}. In the first layer, \textcolor[HTML]{FF8A80}{\texttt{AR762}} predominantly attends to own state information, whereas in the subsequent layers, attention shifts toward other aircraft. In particular, \textcolor[HTML]{EA4335}{\texttt{SIA608}} receives substantially higher attention in both the second and third layers. This interaction pattern is more apparent in the structured traffic flow, where aircraft with similar arrival sequence and close spatial proximity receive greater attention. Although more extensive validation is necessary, these two examples, along with similar observations in other cases, suggest that the attention scores of the model can serve as a valuable tool for interpreting prediction outcomes and supporting ATCo in real-time operations.

\section{Conclusion}\label{conclusion}
In this work, we proposed a probabilistic multi-agent aircraft landing time prediction framework that estimates the landing-time distributions of multiple aircraft. The proposed model combines a multi-agent trajectory encoder with a Gaussian parameter decoder, which outputs both the predicted landing times and their associated uncertainties. Using the ADS-B trajectory dataset collected from the terminal airspace of Incheon International Airport in South Korea, we demonstrated that our model outperformed the baseline models in terms of prediction accuracy and arrival sequence consistency. The predicted uncertainties effectively captured the inherent variability in aircraft landing times, thereby increasing the practical value of the prediction results. Moreover, the agent attention scores provided by the model can be used to interpret its prediction outcomes.

\textbf{Limitations and Future Works.} Despite the promising results, several limitations should be addressed in future research. First, our model was developed solely using trajectory data. Incorporating additional factors, such as weather information, may further improve the predictive performance, because landing times are often influenced by environmental conditions. Second, we assume that the predicted landing times follow a unimodal Gaussian distribution. However, in practice, the landing time distribution may exhibit multimodal or asymmetric characteristics that cannot be fully captured under this assumption. Exploring more flexible distributional representations may enhance both the accuracy and robustness. Finally, the agent attention scores should be validated against human air traffic controllers’ attention patterns, such as those measured through eye-tracking experiments, to confirm their alignment. Such a validation would further strengthen the model’s ability to support human air traffic controllers in real-world operations.

\section*{Acknowledgments}
This work is supported by the Korea Agency for Infrastructure Technology Advancement (KAIA) grant funded by the Ministry of Land, Infrastructure and Transport (Grant RS-2022-00156364).

\bibliography{cite}

\end{document}